%% file: main.tex
\documentclass[conference]{IEEEtran}
\usepackage{cite}
\usepackage{amsmath,amssymb,amsfonts}
\usepackage{algorithmic}
\usepackage{graphicx}
\usepackage{textcomp}
\usepackage[dvipsnames]{xcolor}
\usepackage{colortbl}
\usepackage[linesnumbered,ruled,vlined]{algorithm2e}
\usepackage{caption}
\usepackage{subcaption}
\usepackage{soul}
\usepackage[export]{adjustbox}
\usepackage{amsthm}
\usepackage{comment}
\usepackage{booktabs}
\usepackage{todonotes}

\theoremstyle{definition}

\theoremstyle{definition}

\usepackage{mathtools}

\usepackage{tikz} \newcommand*\circled[1]{\tikz[baseline=(char.base)]{             \node[shape=circle,fill,inner sep=1pt] (char) {\textcolor{white}{#1}};}}

\def\BibTeX{{\rm B\kern-.05em{\sc i\kern-.025em b}\kern-.08em
    T\kern-.1667em\lower.7ex\hbox{E}\kern-.125emX}}

\def\BibTeX{{\rm B\kern-.05em{\sc i\kern-.025em b}\kern-.08em
    T\kern-.1667em\lower.7ex\hbox{E}\kern-.125emX}}

\newcommand{\projname}{Ba-ZebraConf}

\begin{document}  
\title{\projname{}: A Three-Dimension Bayesian Framework for Efficient System Troubleshooting}

\author{
\IEEEauthorblockN{1\textsuperscript{st} Deyi Xing}
\IEEEauthorblockA{\textit{Department of Computer Science and Engineering} \\
\textit{University of California Merced}\\
Merced, USA \\
dxing@ucmerced.edu}
\and
\IEEEauthorblockN{2\textsuperscript{nd} Weicong Chen}
\IEEEauthorblockA{\textit{Department of Computer Science and Engineering} \\
\textit{University of California Merced}\\
Merced, USA \\
wchen97@ucmerced.edu}
\hspace{60.0ex}
\and
\IEEEauthorblockN{3\textsuperscript{rd} Curtis Tatsuoka}
\IEEEauthorblockA{\textit{Department of Medicine} \\
\textit{University of Pittsburgh}\\
Pittsburgh, USA \\
cut4@pitt.edu}
\hspace{55.0ex}
\and
\IEEEauthorblockN{4\textsuperscript{th} Xiaoyi Lu}
\IEEEauthorblockA{\textit{Department of Computer Science and Engineering} \\
\textit{University of California Merced}\\
Merced, USA \\
xiaoyi.lu@ucmerced.edu}
}

\maketitle
\thispagestyle{plain}
\pagestyle{plain}

\begin{abstract}

The proliferation of heterogeneous configurations in distributed systems presents critical challenges in ensuring stability and operational efficiency. Misconfigurations, especially those arising from complex parameter interdependencies, can lead to severe system failures. Group Testing (GT), originally developed to efficiently identify defective items within large populations, can significantly expedite the troubleshooting process by reducing the number of tests compared to exhaustive individual testing. Existing methods, such as ZebraConf, employ a binary-splitting-based GT strategy, demonstrating the feasibility of leveraging GT for large-scale distributed systems. However, this approach suffers from notable limitations, including reliance on sequential testing stages, inability to account for parameter interdependencies, and vulnerability to testing errors such as noise and dilution effects.
To address these challenges, we propose \textbf{\textbf{}}, a novel three-dimensional Bayesian framework that integrates three complementary Bayesian methodologies to redefine heterogeneous configuration testing. First, Bayesian Group Testing (BGT) replaces binary splitting with a probabilistic lattice model, dynamically refining testing strategies through the Bayesian Halving Algorithm (BHA) to prioritize high-risk configurations and adapt to noisy outcomes. Second, Bayesian Optimization (BO) automates the tuning of critical hyperparameters, such as pool sizes and classification thresholds, using a Gaussian Process-based surrogate model and acquisition functions to optimize resource allocation and test efficiency. Finally, we introduce Bayesian Risk Refinement (BRR), which iteratively updates each parameter’s risk assessment across multiple group tests. BRR accumulates evidence through Bayesian inference, automatically capturing interdependencies between parameters and enabling more confident classifications. By progressively refining risk assessments, BRR accelerates troubleshooting and eliminates redundant tests, improving both efficiency and accuracy.
By leveraging these three Bayesian dimensions, \projname{} captures parameter interdependencies, adapts dynamically to noisy environments, and scales efficiently to large configuration spaces. Evaluation results demonstrate that \projname{} reduces the average test count and test execution time by  67\% compared to the original ZebraConf, while decreasing false positive and false negative rates to 0\%, respectively. These improvements establish \projname{} as a robust and scalable solution for dynamic, heterogeneous distributed systems, setting a new benchmark for efficient and accurate system troubleshooting.

\end{abstract}


\input{texts/intro}
\input{texts/background}
\input{texts/problem-statement}
\input{texts/design}
\input{texts/implementation}
\input{texts/evaluation}
\input{texts/conclusion}
\newpage


\bibliographystyle{IEEEtran}
\bibliography{main}


\end{document}

%% file: texts/intro.tex
\section{Introduction}

Distributed systems form the backbone of modern computational infrastructures, powering cloud platforms, edge computing, and IoT ecosystems. These systems are becoming increasingly complex, characterized by heterogeneous configurations involving diverse hardware, software, and network parameters \cite{cheng2016adaptive, lama2012aroma, zhang2019cloudtuning,hdfs2202,mahgoub2020optimuscloud}. While such configurability enables flexibility and optimization, it also introduces significant risks of misconfigurations—errors in parameter settings that degrade performance or cause catastrophic failures \cite{ma2021heterogeneous}. To address this issue,   various methods incrementally modify the configuration of a subset of nodes, either by rebooting these nodes \cite{borthakur2011hadoop, clouderaRollingRestart, mahgoub2019sophia, mysqlRollingRestart, mapreduce442} or by leveraging application APIs \cite{cheng2016adaptive, hbase8544, hdfs1477, li2014mronline, wang2018autoadjust} until the new configuration is applied to all nodes. Efficient diagnosis and resolution of these misconfigurations is critical to maintaining system stability and operational efficiency, but it remains a formidable challenge given the scale and heterogeneity of these systems.

Traditional approaches to troubleshooting \cite{attariyan2010automating, ma2021heterogeneous}, such as exhaustive testing, rule-based diagnostics, and static analysis, struggle to keep pace with the complexity and scale of modern distributed systems. Exhaustive testing is computationally prohibitive due to the exponential growth of configuration spaces, while static analysis and rule-based methods lack robustness in dynamic and noisy environments. These limitations necessitate new methodologies that combine efficiency, adaptability, and robustness to meet the demands of large-scale distributed systems.

Group Testing (GT), originally developed for efficiently identifying defective items in large populations \cite{lohse2020pooling, majid2020optimising, hogan2020sample, donoho2020mathematics}, offers a promising alternative for scalable configuration troubleshooting. GT reduces the number of required tests by pooling parameters and evaluating them simultaneously \cite{chen2023sbgt, lohse2020pooling}. ZebraConf \cite{ma2021heterogeneous}, a seminal framework, adapted GT using a binary-splitting strategy to isolate problematic configurations. While this approach demonstrated the feasibility of using GT for configuration troubleshooting, it suffers from several critical limitations.

\subsection{Motivation and Challenges}

The limitations of ZebraConf highlight the need for an advanced troubleshooting framework. Specifically:

\noindent \textbf{Scalability Issues:} ZebraConf’s binary-splitting strategy requires multiple sequential testing stages, resulting in significant delays as configuration spaces grow.

\noindent \textbf{Noise Susceptibility:} Real-world testing environments introduce noise, leading to dilution effects and inaccuracies, which increase false positives and negatives.

\noindent \textbf{Interdependencies Among Parameters:} 
In heterogeneous systems, parameter interdependencies \cite{chen2020configuration} significantly complicate troubleshooting. For instance, the configuration of one parameter (e.g., \(A\)) can effectively mask the faultiness of another parameter (e.g., \(B\)) in certain group tests. This behavior, where parameter interactions influence unit test outcomes, can cause group testing strategies to miss true faulty parameters or falsely classify safe ones. Resolving these interdependencies requires a systematic mechanism for accumulating evidence across multiple tests.

To address these shortcomings, we introduce \textbf{\projname{}}, a novel framework that integrates three Bayesian methodologies—Bayesian Group Testing (BGT), Bayesian Optimization (BO), and Bayesian Risk Refinement (BRR)—into a unified system for efficient, accurate, and scalable troubleshooting.

Designing an advanced framework like \projname{} requires addressing several challenges:

\noindent \textbf{Challenge 1: Dynamic Test Prioritization.} Identifying and focusing on high-risk configurations in real time is essential for efficiency and accuracy in large configuration spaces.

\noindent \textbf{Challenge 2: Noise Resilience.} Robust methodologies are needed to mitigate noise, dilution effects, and errors in testing outcomes.

\noindent \textbf{Challenge 3: Parameter Interdependencies.} Many configurations exhibit parameter interdependencies that influence outcomes. For example, the faulty behavior of one parameter may be masked by specific settings of another, requiring evidence to be accumulated across multiple tests to make confident classifications.

\noindent \textbf{Challenge 4: Hyperparameter Optimization.} Pool sizes, thresholds, and resource allocations must be dynamically adapted to varying workloads and system conditions.

\noindent \textbf{Challenge 5: Scalability.} The framework must handle the exponential growth of configuration spaces while maintaining computational efficiency.

\subsection{Contributions}

To overcome these challenges, we propose \textbf{\projname{}}, a groundbreaking three-dimensional Bayesian framework that systematically integrates BGT, BO, and BRR. Each Bayesian methodology addresses specific challenges while synergistically enhancing the overall framework:

\noindent \textbf{1. Bayesian Group Testing (BGT):} 
BGT replaces ZebraConf’s binary-splitting strategy with a probabilistic lattice model, using the Bayesian Halving Algorithm (BHA) \cite{chen2023sbgt} to dynamically refine testing strategies. By iteratively updating posterior probabilities based on test outcomes, BGT prioritizes high-risk configurations and mitigates false outcomes, even in noisy environments.

\noindent \textbf{2. Bayesian Optimization (BO):} 
BO automates the tuning of hyperparameters such as pool sizes, prior probabilities, and thresholds, using a Gaussian Process-based surrogate model to efficiently explore the search space \cite{frazier2018tutorial}. The Expected Improvement acquisition function balances exploration and exploitation, ensuring resource allocation and test strategies are continuously optimized.

\noindent \textbf{3. Bayesian Risk Refinement (BRR):} 
BRR addresses the challenge of parameter interdependencies by iteratively updating each parameter’s risk assessment across multiple group tests. Rather than treating individual test results as definitive, BRR accumulates evidence using Bayesian inference, gradually transitioning parameters from prior risk estimates to posterior probabilities. This approach allows the framework to identify hidden faulty parameters that might otherwise be masked in certain test scenarios, while reducing unnecessary testing once a parameter is classified as highly likely to be positive or negative. BRR not only improves classification accuracy but also accelerates troubleshooting by eliminating redundant tests.

\textbf{Key Contributions:}
\noindent \circled{1} \textbf{Three-Dimensional Bayesian Framework:} We introduce the first seamless integration of BGT, BO, and BRR into a unified troubleshooting system for distributed configurations.
\noindent \circled{2} \textbf{Dynamic and Noise-Resilient Testing:} BGT and BRR mitigate noise and dynamically refine testing strategies to minimize false outcomes.
\noindent \circled{3} \textbf{Automated Hyperparameter Optimization:} BO automates the tuning of critical parameters, improving adaptability and efficiency in diverse environments.
\noindent \circled{4} \textbf{Holistic Interdependency Modeling:} BRR accumulates evidence across tests, capturing complex parameter interactions that traditional methods overlook.
\noindent \circled{5} \textbf{Scalable and Distributed Framework:} Optimized pooling strategies and distributed computation enable efficient handling of large-scale configuration spaces.

\textbf{To the best of our knowledge}, \projname{} represents the first seamless integration of Bayesian Group Testing, Bayesian Optimization, and Bayesian Risk Refinement for configuration troubleshooting. The probabilistic foundation shared by these three methodologies enables:
\textbf{(1)} Adaptive and noise-resilient test prioritization,
\textbf{(2)} Dynamic optimization of resource allocation and test strategies,
and \textbf{(3)} Scalable and efficient handling of large configuration spaces.

This work establishes a new benchmark for scalable and robust system troubleshooting, offering transformative improvements in efficiency, accuracy, and adaptability for modern distributed systems.

%% file: texts/background.tex
\section{Background}

This section introduces the foundational elements of \projname{}: ZebraConf for identifying heterogeneous-unsafe parameters, Bayesian Group Testing for probabilistic adaptive testing, and Bayesian Optimization for hyperparameter tuning. These methods form the building blocks of a scalable, robust, and efficient troubleshooting framework for distributed systems.

\subsection{Cloud System Troubleshooting by Finding Heterogeneous-Unsafe Parameters}

Distributed systems often assume homogeneous configurations across all nodes \cite{cheng2016adaptive, lama2012aroma, zhang2019cloudtuning,hdfs2202,mahgoub2020optimuscloud, ma2021heterogeneous}. However, heterogeneous configurations are increasingly adopted for two key reasons. First, heterogeneous hardware necessitates tailored configurations to achieve optimal performance. Second, even in homogeneous systems, runtime configuration changes are often required to adapt to varying workloads without rebooting the entire system, which can be disruptive  \cite{borthakur2011hadoop, clouderaRollingRestart, mahgoub2019sophia, mysqlRollingRestart, mapreduce442, ma2021heterogeneous}. 

Despite these benefits, improperly implemented heterogeneous configurations can cause system failures \cite{ma2021heterogeneous}. These failures often stem from complex parameter interdependencies, where one parameter’s configuration can mask or amplify another’s faultiness. For example, setting parameter \(A\) to a specific range might hide the faultiness of parameter \(B\), leading to failures when \(A\) is adjusted. These complexities make it challenging to identify true heterogeneous-unsafe parameters, particularly in noisy environments.

ZebraConf is a state-of-the-art framework for identifying heterogeneous-unsafe parameters \cite{ma2021heterogeneous}. It operates by systematically running unit tests on both homogeneous and heterogeneous configurations. ZebraConf consists of three components: \textbf{TestGenerator}, which selects unit tests and determines configurations; \textbf{TestRunner}, which executes tests and identifies errors caused by heterogeneous configurations; and \textbf{ConfAgent}, which manages the execution of tests. While effective, ZebraConf faces challenges in scalability, handling noisy environments, and accounting for parameter interdependencies.

\subsection{Bayesian Group Testing (BGT)}

BGT combines the principles of traditional group testing with Bayesian inference to efficiently classify a population while accounting for uncertainties such as noise and dilution effects \cite{chen2023sbgt,tatsuoka2022bayesian}. Traditional group testing, as introduced by Dorfman, involves pooling multiple items and testing them collectively. A negative result clears all items in the pool, while a positive result necessitates further subdivision to isolate defective items. BGT enhances this approach by incorporating probabilistic reasoning and adaptive test selection, guided by a Bayesian framework \cite{chen2023sbgt, tatsuoka2022bayesian}.

The central idea of BGT is to iteratively update the posterior probability distribution of a configuration space based on test outcomes \cite{chen2023sbgt}. Let $\mathbf{x} = \{x_1, x_2, \ldots, x_N\}$ represent the states of $N$ items, where $x_i \in \{0, 1\}$ indicates whether item $i$ is defective. A pooled test evaluates whether any item in the group is defective:
\[
y = 
\begin{cases} 
1 & \text{if } \bigcup_{i \in G} x_i = 1, \\
0 & \text{otherwise,}
\end{cases}
\]
where $G$ is the set of indices in the group. The likelihood of observing test outcome $y$ given state $\mathbf{x}$ is:
\[
P(y \mid \mathbf{x}) = 
\begin{cases} 
1 - \prod_{i \in G} (1 - x_i) & \text{if } y = 1, \\
\prod_{i \in G} (1 - x_i) & \text{if } y = 0.
\end{cases}
\]

Using Bayes' theorem, the posterior probability of a configuration $\mathbf{x}$ after observing outcomes $\mathbf{y}$ is:
\[
P(\mathbf{x} \mid \mathbf{y}) = \frac{P(\mathbf{y} \mid \mathbf{x}) P(\mathbf{x})}{P(\mathbf{y})},
\]
where $P(\mathbf{x})$ is the prior probability of $\mathbf{x}$, and $P(\mathbf{y})$ normalizes the distribution.

BGT leverages lattice-based models \cite{tatsuoka2022bayesian, tatsuoka2003sequential, chen2023sbgt} to efficiently represent and manipulate the configuration space. The lattice structure $(L, \leq)$ captures all possible diagnostic outcomes, with $\leq$ representing the subset relationship. Each node in the lattice corresponds to a subset of parameters, and its position reflects its likelihood of containing defective elements.

The Bayesian Halving Algorithm (BHA) \cite{chen2023sbgt} selects tests to maximally reduce uncertainty by partitioning the lattice into \textit{two nearly equal parts based on posterior probabilities}. The test selection process aims to minimize the expected posterior entropy across the lattice:
\[
\mathbb{E}[H_{\text{posterior}}] = -\sum_{\mathbf{x}} P(\mathbf{x} \mid \mathbf{y}) \log P(\mathbf{x} \mid \mathbf{y}).
\]
The BHA iteratively identifies the most informative tests, updating the posterior probabilities after each test to refine the search space \cite{chen2023sbgt}.

A critical advantage of BGT is its ability to model and mitigate noise and dilution effects \cite{chen2023sbgt, tatsuoka2022bayesian}, which are common in real-world pooled testing scenarios. For example, dilution effects occur when the inclusion of non-defective items in a pool masks defective ones, leading to false negatives. BGT incorporates these effects into the likelihood models, ensuring robust performance.

Scaling Bayesian-based Group Testing (SBGT) \cite{chen2023sbgt} further optimizes BGT by introducing bitwise operations for lattice manipulation, significantly reducing computational complexity. By leveraging order-theoretic properties of the lattice, SBGT enables efficient test selection and posterior updates, making it suitable for large-scale configuration spaces.

\subsection{Bayesian Optimization (BO)}

BO is a probabilistic framework for optimizing expensive black-box functions \cite{frazier2018tutorial, NIPS2012_05311655,10.1145/3545611, berk2019exploration, 10.1145/3582078}. Given an objective function $f: \mathcal{X} \to \mathbb{R}$, where $\mathcal{X}$ is the search space, the goal is to find $x^* = \arg \max_{x \in \mathcal{X}} f(x)$ with minimal evaluations of $f$. The process involves three key components:
\begin{itemize}
    \item \textbf{Surrogate Model:} A probabilistic model approximates the objective function $f$. Gaussian Processes (GPs) are commonly used \cite{NIPS2012_05311655,10.1145/3545611}, where the posterior distribution over $f(x)$ is given by:
    \[
    f(x) \sim \mathcal{GP}\left(m(x), k(x, x')\right),
    \]
    with mean function $m(x)$ and kernel function $k(x, x')$ representing covariance between points.
    \item \textbf{Acquisition Function:} This function, $\alpha(x)$, balances exploration (sampling uncertain regions) and exploitation (focusing on promising regions). A popular choice is Expected Improvement (EI) \cite{berk2019exploration,10.1145/3582078}:
    \[
    \alpha(x) = \mathbb{E}\left[\max(f(x) - f_{\text{best}}, 0)\right],
    \]
    where $f_{\text{best}}$ is the best observed value so far.
    \item \textbf{Sampling Strategy:} BO iteratively updates the surrogate model and uses $\alpha(x)$ to determine the next point $x_{n+1}$ for evaluation.
\end{itemize}

The BO process begins with an initial sampling of $f$ to construct the surrogate model. At each iteration, the acquisition function selects the next evaluation point, and the surrogate model is updated with the new observation. This iterative optimization continues until convergence or a predefined budget is reached \cite{frazier2018tutorial}. BO is particularly useful for tuning hyperparameters in BGT, such as pool sizes or thresholds, ensuring efficient resource allocation.

%% file: texts/problem-statement.tex
\section{Problem Statement and Objectives}

Modern distributed systems increasingly rely on heterogeneous configurations to optimize performance and adapt to dynamic workloads. However, these configurations pose significant risks of misconfigurations, particularly when parameter interdependencies mask or amplify faults, leading to catastrophic failures such as communication breakdowns or degraded system performance. Identifying these heterogeneous-unsafe parameters is essential for ensuring system reliability, yet it remains a challenging task in dynamic and large-scale environments.

Existing frameworks like ZebraConf \cite{ma2021heterogeneous} address this issue using group testing strategies but face several limitations:
\begin{itemize}
    \item \textbf{Scalability Issues:} Sequential testing in binary-splitting strategies leads to delays as configuration spaces grow.
    \item \textbf{Noise Susceptibility:} ZebraConf lacks mechanisms to mitigate noise and dilution effects, resulting in false positives and negatives.
    \item \textbf{Limited Interdependency Modeling:} The framework assumes parameter independence, missing faults arising from interdependent configurations. For instance, certain parameter combinations may hide the faultiness of individual parameters, leading to incomplete or incorrect classifications.
\end{itemize}

Addressing these limitations requires a scalable and robust solution that can:
\begin{itemize}
    \item Dynamically prioritize high-risk configurations to improve testing efficiency.
    \item Accurately model and address parameter interdependencies to minimize classification errors.
    \item Mitigate noise and ensure reliable detection in real-world environments.
    \item Terminate testing for parameters once a high degree of confidence is reached, improving overall efficiency.
\end{itemize}

\textbf{Formal Problem Definition.} To address these challenges, we develop \textbf{\projname{}}, a framework that integrates Bayesian Group Testing (BGT), Bayesian Optimization (BO), and Bayesian Risk Refinement (BRR) to minimize computational costs while maximizing detection accuracy. The problem can be formalized as follows:
\begin{itemize}
    \item \(C\): Computational cost of classification, representing the total cost of performing tests and analyzing results.
    \item \(A\): Classification accuracy, representing the correctness of identifying heterogeneous-unsafe parameters.
    \item \(\theta_i\): Hyperparameters for the Bayesian test selector, controlling the test selection and optimization process.
    \item \(F_i\): The \(i\)-th configuration file in a set of \(n\) files, each defining the system's configuration.
    \item \(\text{Conf}(F_i)\): The heterogeneous parameter configuration derived from the configuration file \(F_i\).
    \item \(\lambda\): A weight parameter balancing the trade-off between computational cost and accuracy.
\end{itemize}

The objective is to minimize the weighted sum of computational costs and accuracy penalties over all configurations:
\[
\min \sum_{i=1}^n \left( C(\theta_i, \text{Conf}(F_i)) + \lambda \cdot (1 - A(\theta_i, \text{Conf}(F_i))) \right),
\]
where \(C(\theta_i, \text{Conf}(F_i))\) models the resource consumption of testing under the given hyperparameters and configurations, and \(A(\theta_i, \text{Conf}(F_i))\) quantifies the detection accuracy. The term \((1 - A)\) ensures that low accuracy incurs a penalty, incentivizing reliable detection.

\textbf{Design Objectives.} To solve this problem, \projname{} introduces:
\begin{itemize}
    \item \textbf{Probabilistic Prioritization:} BGT dynamically prioritizes high-risk configurations, reducing redundant tests while maintaining accuracy.
    \item \textbf{Adaptive Hyperparameter Optimization:} BO tunes critical parameters such as pool sizes and thresholds to balance exploration and exploitation, ensuring efficient resource allocation under diverse conditions.
    \item \textbf{Interdependency-Aware Testing:} BRR leverages Bayesian inference to refine the risk assessment of each parameter over multiple tests. This refinement process accounts for parameter interdependencies and eliminates redundant testing by terminating further evaluations once a parameter is deemed conclusively faulty or safe.
    \item \textbf{Noise Resilience:} BGT models testing errors, such as noise and dilution effects, enabling robust classification even in imperfect testing environments.
\end{itemize}

%% file: texts/design.tex
\section{Design}

\projname{} introduces a transformative approach to identifying heterogeneous-unsafe parameters in distributed systems. By integrating BGT, BRR, and BO, \projname{} overcomes the limitations of traditional methods, such as inefficiency, inaccuracy, and poor scalability. This novel three-dimensional Bayesian framework leverages probabilistic modeling, iterative refinement, and dynamic hyperparameter tuning to enhance troubleshooting capabilities in complex system configurations.

\begin{figure}[h]
    \centering
    \includegraphics[width=1\linewidth]{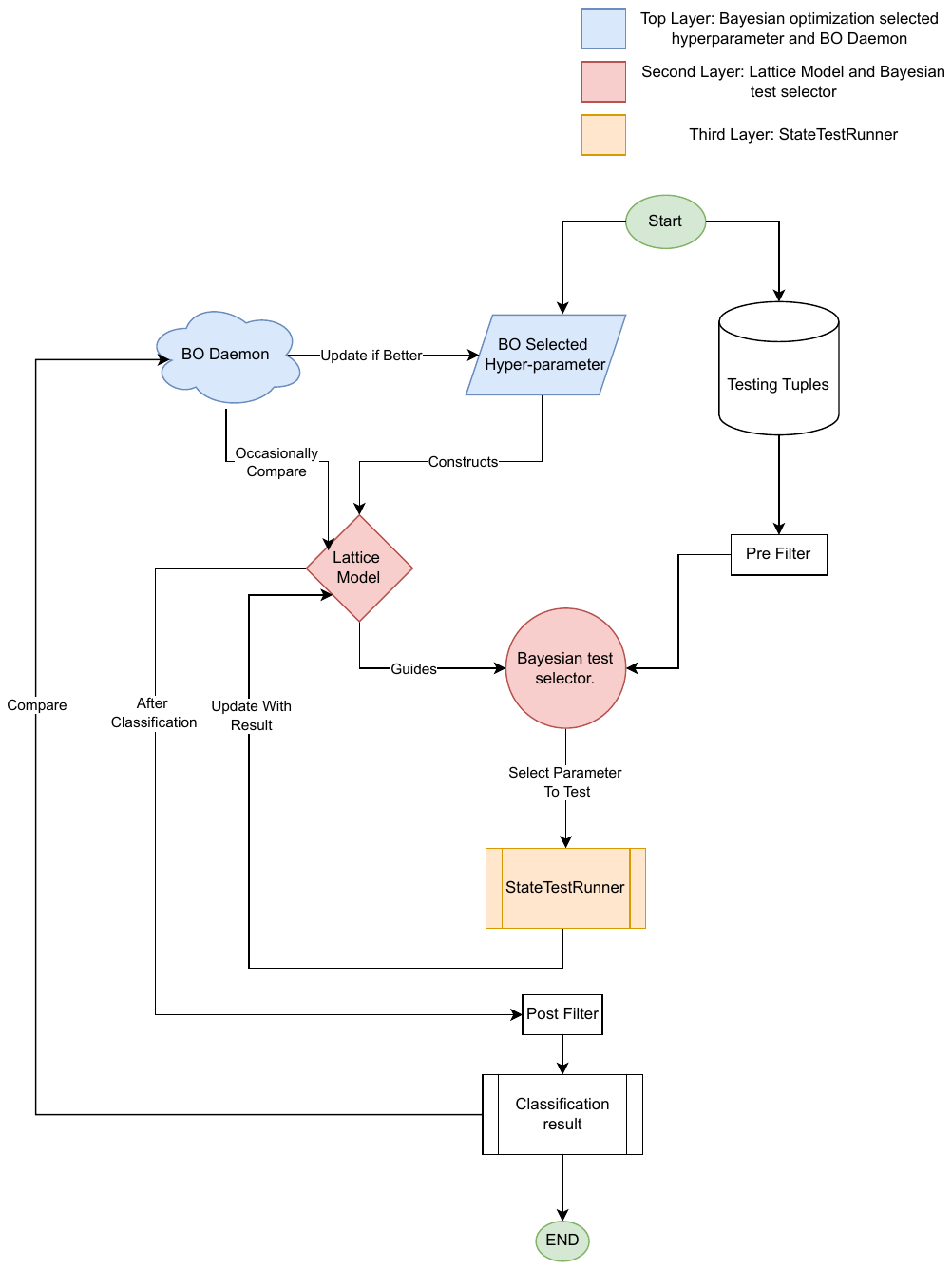}
    \caption{Overview of \projname{}. The framework incorporates Bayesian Optimization for adaptive hyperparameter tuning, Bayesian Group Testing for probabilistic inference, and Bayesian Risk Refinement for cumulative risk assessment and interdependency modeling.}
    \label{fig:overview}
\end{figure}

\subsection{Overview}

As illustrated in Figure~\ref{fig:overview}, \projname{} consists of three synergistic modules that collectively enhance the efficiency and accuracy of the testing process:
\begin{itemize}
    \item \textbf{Bayesian Group Testing Module:} Constructs probabilistic lattice models to prioritize and test high-risk parameter configurations. This module addresses parameter dependencies, mitigates dilution effects, and dynamically updates posterior probabilities.
    \item \textbf{Bayesian Risk Refinement Module:} Iteratively updates each parameter's risk assessment by combining the results of multiple tests using Bayesian inference. It enables cumulative risk evaluation and the dynamic termination of tests for parameters classified as definitively safe or unsafe.
    \item \textbf{Bayesian Optimization Module:} Dynamically tunes hyperparameters to adapt testing strategies to varying workloads and system behaviors. This module ensures efficient allocation of resources and improved testing outcomes through continuous refinement.
\end{itemize}

This unified framework synergistically combines the strengths of BGT, BRR, and BO, enabling adaptive, probabilistic, and efficient configuration troubleshooting in large-scale distributed systems.

\subsection{Bayesian Group Testing Module}

The BGT module constructs a probabilistic lattice model $(L, \leq)$, where:
\begin{itemize}
    \item Each node represents a subset of parameters with associated probabilities of being faulty.
    \item Directed edges represent subset relationships, facilitating efficient traversal and updates.
\end{itemize}

The lattice model is initialized using prior probabilities for each parameter, which reflect expert knowledge about their likelihood of being erroneous. These priors provide a foundational probabilistic mapping that evolves dynamically as test results become available. After each unit test execution, the Bayesian test selector updates the lattice model to refine the posterior probabilities of parameter faultiness. This iterative updating process ensures that the system continuously adapts to new observations, focusing resources on the most uncertain or high-risk configurations.

\textbf{Key Designs in the BGT Module:}

\subsubsection{Global Prior Probability Mapping for Parameter Classification}

To capture the classification history and ensure a systematic evaluation of heterogeneous-unsafe parameters, the BGT module introduces a global prior probability mapping for each parameter. Initially, each parameter is assigned a prior probability based on expert understanding of its potential faultiness. As testing progresses:
\begin{itemize}
    \item If a parameter is classified as heterogeneously unsafe, its prior probability for subsequent tests is increased.
    \item Conversely, if a parameter is classified as negative, its prior probability is decreased.
\end{itemize}

This adaptive prior updating accelerates future evaluations and allows the framework to leverage historical test results for more informed classifications. By the end of the full test run, the accumulated prior probabilities enable a holistic assessment of each parameter's behavior, minimizing the reliance on single test outcomes. This approach avoids premature conclusions and enhances overall classification accuracy.

\subsubsection{Handling Parameter Dependencies and Reducing False Positives}

One of the key limitations of ZebraConf’s binary-splitting strategy is its inability to handle parameter interdependencies effectively. Parameter dependencies can lead to false positives or missed faults when configurations are fragmented into subgroups. For example:
\begin{itemize}
    \item A parameter combination \{A, B, C\} might test positive as a group but yield false negatives when tested individually or in smaller subgroups (\{A, B\} and \{C\}, for instance).
    \item Such inconsistent behaviors indicate potential dependencies between parameters, which the binary-splitting approach cannot capture.
\end{itemize}

To address this, the Bayesian test selector tracks the state of each test selection and evaluates inconsistencies. If repeated testing confirms dependency-related inconsistencies, the original combination (\{A, B, C\}) is flagged for further analysis. After a predefined number of iterations, such combinations are classified as exhibiting dependency relationships and are saved for manual interpretation or deeper testing.

This mechanism not only reduces false negatives but also ensures that parameter interdependencies are accounted for, improving both accuracy and system reliability.

\subsubsection{Dilution-Resilient Pooling and Lattice Shrinking}

The BGT module employs adaptive pooling strategies to mitigate the dilution effect, where the presence of multiple non-faulty parameters in a pool can obscure the detection of a faulty one. By dynamically adjusting the size and composition of test pools, the module ensures balanced sensitivity and efficiency in detecting errors. Additionally, as parameters or parameter subsets are classified, they are pruned from the lattice model. This lattice shrinking reduces computational overhead and focuses resources on unresolved configurations, ensuring scalability to large configuration spaces.

\subsection{BRR Module}

BRR enables cumulative risk assessment by iteratively updating each parameter's risk level based on multiple test results. Let $R_i^{(t)}$ denote the risk level of parameter $i$ after $t$ tests. Using Bayesian inference:
\begin{equation*}
P(R_i^{(t)} \mid \text{tests}) = 
\frac{P(\text{tests} \mid R_i^{(t)}) P(R_i^{(t)})}{P(\text{tests})},
\end{equation*}
where $P(R_i^{(t)})$ is the prior risk estimate, and $P(\text{tests} \mid R_i^{(t)})$ is the likelihood of observing the test results given the risk level.

The likelihood function, tuned by BO, determines the weight assigned to positive and negative test results. For example:
\begin{align*}
P(\text{positive test} \mid R_i) &= \alpha R_i, \\
P(\text{negative test} \mid R_i) &= 1 - \beta R_i,
\end{align*}
where $\alpha$ and $\beta$ are parameters controlled by BO to reflect system noise or dilution effects.

Risk thresholds $\tau_{\text{safe}}$ and $\tau_{\text{unsafe}}$ are used to classify parameters:
\begin{align*}
\text{Classify as safe if } & R_i^{(t)} < \tau_{\text{safe}}, \\
\text{Classify as unsafe if } & R_i^{(t)} > \tau_{\text{unsafe}}.
\end{align*}

Key features include:
\begin{itemize}
    \item \textbf{Cumulative Risk Assessment:} Accounts for all observed test results, providing a holistic evaluation of each parameter.
    \item \textbf{Dynamic Test Termination:} Stops testing parameters once their risk levels reach classification thresholds, reducing redundant tests.
    \item \textbf{Interdependency Mitigation:} Incorporates risk adjustments for parameters with dependent behaviors, improving classification accuracy.
\end{itemize}

\subsection{BO Module}

We employ a tailored BO strategy to dynamically navigate the hyperparameter search space to identify optimal configurations tailored for \projname{}'s BGT and BRR modules. By leveraging a Gaussian Process (GP)-based surrogate model and the Expected Improvement (EI) acquisition function, BO efficiently balances exploration and exploitation to improve both testing efficiency and classification accuracy in large-scale distributed systems.

\textbf{Surrogate Modeling and Acquisition.} 
BO models the optimization problem as:
\[
\mathbf{x}^* = \arg \min_{\mathbf{x} \in \mathcal{X}} f(\mathbf{x}),
\]
where $\mathbf{x}$ represents hyperparameters such as pool sizes, prior probabilities, classification thresholds, and likelihood functions for BRR. $\mathcal{X}$ is the hyperparameter space, and $f(\mathbf{x})$ is the objective function representing the trade-off between testing efficiency and accuracy.

The GP surrogate model predicts the objective function’s behavior:
\[
f(\mathbf{x}) \sim \mathcal{GP}(m(\mathbf{x}), k(\mathbf{x}, \mathbf{x'})),
\]
where $m(\mathbf{x})$ is the mean function and $k(\mathbf{x}, \mathbf{x'})$ is the kernel function defining covariance. The EI acquisition function is used to select the next set of hyperparameters for evaluation:
\[
\alpha_{\text{EI}}(\mathbf{x}) = \mathbb{E}\left[\max(f_{\text{best}} - f(\mathbf{x}), 0)\right],
\]
where $f_{\text{best}}$ is the best observed value so far.

\textbf{Two-Phase Optimization Process.} 
The BO module separates the optimization process into two distinct phases: the Exploration Phase and the Exploitation Phase, each tailored to balance the trade-offs between comprehensive exploration and efficient exploitation of promising hyperparameter regions.

\subsubsection{Exploration Phase}
In the initial Exploration Phase, the first 200 unit test files are used to run Bayesian optimization continuously. During each iteration, a hyperparameter combination is suggested by the GP model and evaluated on test files sampled systematically to avoid bias. For instance, files from positions $i \cdot 10$ to $(i + 1) \cdot 10$ in the test set are selected. This ensures a representative coverage of the configuration space while building a robust surrogate model that accurately reflects the behavior of the entire test set.

\subsubsection{Exploitation Phase with Bayesian Optimization Daemon}
Following the initial 200 test files, the Exploitation Phase focuses on refining hyperparameters based on the surrogate model built during the Exploration Phase. The best hyperparameter combination identified thus far is used to test the remaining unit files. A Bayesian Optimization Daemon operates in the background during this phase, proposing new hyperparameter combinations based on updated data. Occasionally, subsets of 10 unit files are selected to test these new combinations. If a proposed combination demonstrates superior performance, the Bayesian test selector adopts it for subsequent tests. This adaptive refinement ensures that the framework exploits the best-known configurations while continuing to explore potentially better hyperparameter settings.

\textbf{Tuning Likelihood Functions in BRR.} 
A key novelty of the BO module lies in its ability to dynamically tune the likelihood functions used in BRR. For example, the parameters $\alpha$ and $\beta$ in the likelihood equations:
\[
P(\text{positive test} \mid R_i) = \alpha R_i, \quad P(\text{negative test} \mid R_i) = 1 - \beta R_i,
\]
are adjusted to reflect system conditions such as noise or dilution effects. If false negatives are prevalent, BO increases the weight assigned to positive classifications, accelerating convergence and improving classification accuracy. This synergistic interplay between BO and BRR exemplifies the three-dimensional Bayesian framework’s adaptability and robustness.

\textbf{Automation of Hyperparameter Selection.}
Through this two-phase process, the BO module automates the hyperparameter selection process, optimizing testing strategies for varying workloads and system conditions. This automation reduces the need for manual tuning, enabling \projname{} to scale effectively to large and dynamic configuration spaces.

\textbf{Key Advantages of BO Integration:}
\begin{itemize}
    \item \textbf{Adaptive Optimization:} Continuous refinement ensures optimal testing strategies under dynamic system conditions.
    \item \textbf{Efficient Resource Allocation:} Balances exploration and exploitation to minimize testing time while maximizing accuracy.
    \item \textbf{Dynamic Response to System Noise:} Tuning of BRR’s likelihood functions mitigates noise and dilution effects, improving overall robustness.
\end{itemize}

\subsection{Synergistic Benefits of BGT, BRR, and BO}

The integration of BGT, BRR, and BO creates a three-dimensional Bayesian framework with synergistic benefits:
\begin{itemize}
    \item \textbf{Dynamic Adaptation:} BO dynamically tunes hyperparameters for both BGT and BRR, ensuring optimal performance.
    \item \textbf{Probabilistic Precision:} BGT and BRR refine risk assessments iteratively, capturing interdependencies and mitigating noise effects.
    \item \textbf{Scalable Efficiency:} Lattice shrinking and dynamic test termination reduce computational overhead, enabling scalability to large configuration spaces.
\end{itemize}

\subsection{Test Execution Module}

The Test Execution module translates probabilistic strategies into concrete tests:
\begin{itemize}
    \item \textbf{StateTestRunner:} Converts lattice states into parameter configurations for testing.
    \item \textbf{Controlled Execution:} Executes tests in isolated environments, such as containers, to prevent interference.
    \item \textbf{Result Interpretation:} Analyzes test outcomes to update posterior probabilities in the BGT and BRR modules.
    \item \textbf{Feedback Loop:} Iteratively refines testing strategies based on updated probabilities.
\end{itemize}

This module ensures efficient and accurate testing, with mechanisms to handle inconclusive or conflicting results.

%% file: texts/implementation.tex
\section{Implementation}

The implementation of \projname{} involved significant enhancements to the existing ZebraConf framework, integrating advanced Bayesian methods while maintaining efficiency and scalability.

\subsection{Integration with ZebraConf}

To integrate heterogeneous configuration testing with BGT, we implemented the process outlined in Algorithm~\ref{alg:bgt-Hconf-test}. The integration required:
\begin{itemize}
    \item \textbf{Probabilistic Priors:} We initialized the testing framework with prior probabilities based on domain knowledge, reflecting the likelihood of each parameter being heterogeneously unsafe.
    \item \textbf{Lattice Model Construction:} Using the prior probabilities and hyperparameters optimized through Bayesian methods, we constructed a lattice model representing the configuration space. Each lattice node corresponds to a subset of parameters, with edges encoding subset relationships.
    \item \textbf{Feedback Integration:} A posterior probability feedback loop dynamically updates the lattice model after each test result, enabling iterative refinement of the testing strategy.
    \item \textbf{Backup and Recovery Mechanisms:} To address directory corruption issues caused by invalid configurations, we implemented snapshot-based recovery. The system creates backups of the test environment before each test and restores it post-execution to ensure a clean state for subsequent tests.
    \item \textbf{Filtering Mechanisms:} Parameters frequently causing invalid test results (unrelated to heterogeneous unsafe configurations) are flagged and excluded from the test set. This minimizes redundant tests and improves resource allocation.
\end{itemize}

\SetKwInput{KwInput}{Input}                
\SetKwInput{KwOutput}{Output}             
\SetKwComment{Comment}{$\triangleright$ }{}
\let\oldnl\nl
\newcommand{\nonl}{\renewcommand{\nl}{\let\nl\oldnl}}
\newcommand{\algrule}[1][.8pt]{\par\vskip.25\baselineskip\hrule height #1\par\vskip.25\baselineskip}
\SetKwProg{myproc}{Procedure}{}{}
\SetKwProg{mysubroutine}{Subroutine}{}{}

    


\subsection{Implementation of BGT}

The BGT module systematically identifies heterogeneous unsafe parameters using the probabilistic lattice model and Bayesian test selection strategies. Key implementation details include:

\noindent\textbf{Prior Probability Incorporation:}
Domain knowledge is used to assign higher prior probabilities to parameters deemed high-risk (e.g., encryption settings, resource allocation parameters). During lattice construction, these priors guide the initial allocation of testing resources, allowing the system to focus on configurations more likely to fail. 

\noindent\textbf{In-Place Lattice Shrinking:}
To improve computational efficiency, the lattice model dynamically shrinks as parameters are classified. States corresponding to classified parameters are pruned, and their posterior probabilities are redistributed to remaining states. This optimization reduces computational overhead and accelerates convergence without sacrificing accuracy.

\noindent\textbf{Dependency Handling and False Positive Mitigation:}
We introduced mechanisms to address parameter dependencies, which were a limitation in the original ZebraConf framework. The system tracks test histories and re-evaluates parameter combinations showing inconsistent outcomes. If dependencies are confirmed, the affected parameters are flagged for manual interpretation. This approach mitigates false positives and enhances classification reliability.

\subsection{Implementation of BRR}

The BRR module refines the risk assessment of each parameter iteratively using Bayesian inference. Key implementation details include:

\noindent\textbf{Likelihood Function Tuning:}
The likelihood functions used in BRR are dynamically adjusted to reflect test outcomes. For example:
\[
P(\text{positive test} \mid R_i) = \alpha R_i, \quad P(\text{negative test} \mid R_i) = 1 - \beta R_i,
\]
where $\alpha$ and $\beta$ are tuned via the BO module to account for noise and dilution effects. Parameters classified with high confidence are removed from further testing, reducing redundant evaluations.

\noindent\textbf{Dynamic Risk Thresholding:}
Risk thresholds (\(\tau_{\text{safe}}\) and \(\tau_{\text{unsafe}}\)) are dynamically updated based on test results and system priorities. Parameters are classified as safe or unsafe when their posterior risk levels fall below or above these thresholds, respectively. This dynamic approach accelerates testing convergence while maintaining accuracy.

\subsection{Implementation of BO}

The BO module automates the tuning of six critical hyperparameters for the BGT and BRR modules, including pool sizes, prior probabilities, and classification thresholds. Key implementation details include:

\noindent\textbf{Parallelized Optimization:}
To accelerate convergence, we implemented parallelized hyperparameter evaluations using Docker containers. Each container evaluates 200 candidate configurations simultaneously, leveraging multi-core processing for efficiency.

\noindent\textbf{Two-Phase Optimization Process:}

The first 200 unit tests are used to build a surrogate model of the hyperparameter search space. Bayesian optimization continuously refines this model to identify promising regions. This stage serves as the \textit{exploration phase}.
After this, the best-performing hyperparameters from the Exploration Phase are applied to the remaining tests. A Bayesian Optimization Daemon occasionally explores new hyperparameters to further refine the model. This stage is known as the \textit{exploitation phase}. 

\noindent\textbf{Custom Objective Function:}
Our tailored BO module optimizes hyperparameters using a weighted objective function:
\[
\text{Objective} = 0.4 \cdot \text{TSN} + 0.4 \cdot \text{CSN} - \text{FPR} - \text{FNR},
\]
where TSN and CSN represent time and cost savings, and FPR and FNR are false positive and false negative rates. The weights balance efficiency and accuracy.

\subsection{Engineering Enhancements Over ZebraConf}

To enhance the robustness and scalability of \projname{}, several engineering optimizations were implemented:

\noindent\textbf{Backup and Recovery Mechanisms:} A snapshot system ensures that invalid configurations do not corrupt subsequent tests. Before each test, the environment is saved, allowing for quick restoration in case of corruption. This mechanism significantly reduces downtime and ensures testing reliability, especially when handling configurations prone to failure.

\noindent\textbf{Filtered Testing:} Configurations with a high likelihood of failure due to known issues (e.g., historical failures or incompatibilities) are filtered out early in the process. This reduces unnecessary test executions, conserves resources, and allows the system to focus on more promising configurations. Problematic configurations flagged during testing are marked for further manual analysis.

\noindent\textbf{Logging and Monitoring Tools:} Comprehensive logging captures detailed information about test progress, outcomes, and performance metrics. Real-time monitoring provides insights into system performance, aiding in debugging and validation. This ensures transparency in operations and simplifies troubleshooting.

\noindent\textbf{Isolation of Test Environments:} Each test is executed in an isolated containerized environment, preventing interference from other tests and ensuring that failures or side effects do not propagate across tests. This improves the reliability of test outcomes and maintains system integrity during testing.

\subsection{Validation and Testing}

The effectiveness of \projname{} was validated through extensive testing in both simulated and real-world environments:

\noindent\textbf{Simulated Environments:} Controlled experiments with known failure modes were conducted to verify the system’s ability to identify heterogeneous-unsafe configurations. These tests confirmed the accuracy of \projname{}’s probabilistic classification and its ability to handle noisy environments.

\noindent\textbf{Real-World Deployment:} \projname{} was deployed in live distributed systems with dynamic workloads and diverse configurations. It demonstrated adaptability, successfully identifying unsafe parameters and reducing operational disruptions.

\noindent\textbf{Performance Metrics:} Key performance indicators, including test reduction ratio, time savings, and accuracy rates, were evaluated. For example, in a simulated environment with 100 parameters, \projname{} reduced the number of tests required by 60\% compared to traditional methods, while maintaining a high classification accuracy. This highlights its ability to balance efficiency and robustness effectively.

%% file: texts/evaluation.tex
\section{Evaluation}
\label{sec:evaluation}

This section evaluates the performance of \projname{} across multiple metrics, focusing on execution time, the number of executed tests, and classification accuracy. We compare the results of \projname{} (evaluated at different iterations of Bayesian optimization) against the baseline ZebraConf framework. The analysis demonstrates how \projname{} improves upon ZebraConf by leveraging Bayesian optimization and probabilistic reasoning.

\subsection{Experimental Setup}
The experiments were conducted on a Cloudlab cluster \cite{cloudlab} with the specifications detailed in Table~\ref{tab:spec}. We evaluate Bayesian-ZebraConf at iterations $X = 2, 13, 96, 179$ of the Bayesian optimization process, chosen for their significant performance improvements. Each iteration reflects a refined hyperparameter set derived from Bayesian optimization.

\begin{table}[htbp]
\caption{Specification of Experimental Cluster}
\label{tab:spec}
\centering
\footnotesize
\begin{tabular}{ll}
\toprule
\textbf{Cluster Feature}    & \textbf{Specification}   \\ \hline
Processor     & Two Intel Xeon Silver 4114  \\ 
Clock Speed   & 2.20 GHz  \\
Number of Cores    & 10 per node  \\ 
RAM    & 192GB   \\ 
Storage      & Intel DC S3500 480 GB SATA SSD  \\ 
Scale        & Single-node (10 cores)  \\ 
\bottomrule
\end{tabular}
\end{table}

\noindent Experiments were conducted on 1,000 unit test files to ensure comprehensive evaluation, with metrics including execution time, executed test count, and false positive/negative rates.

\subsection{Evaluation of Execution Time}
\label{ss:lattice-model-benchmark}

\begin{figure}[htbp]
    \centering
    \includegraphics[width=0.80\linewidth]{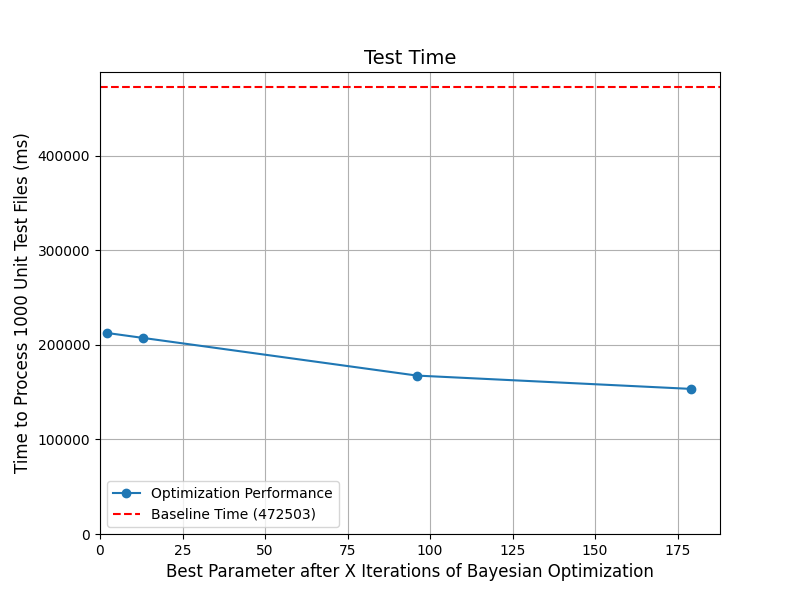}
    \caption{Comparison of Execution Time Between Bayesian ZebraConf (at Iterations $X = 2, 13, 96, 179$) and Original ZebraConf for Evaluating 1,000 Unit Test Files. Bayesian ZebraConf achieved execution times of 212,609s, 207,327s, 167,494s, and 153,417s, compared to ZebraConf's 472,503s.}
    \label{fig:test_time}
\end{figure}

\begin{figure}[htbp]
    \centering
    \includegraphics[width=0.80\linewidth]{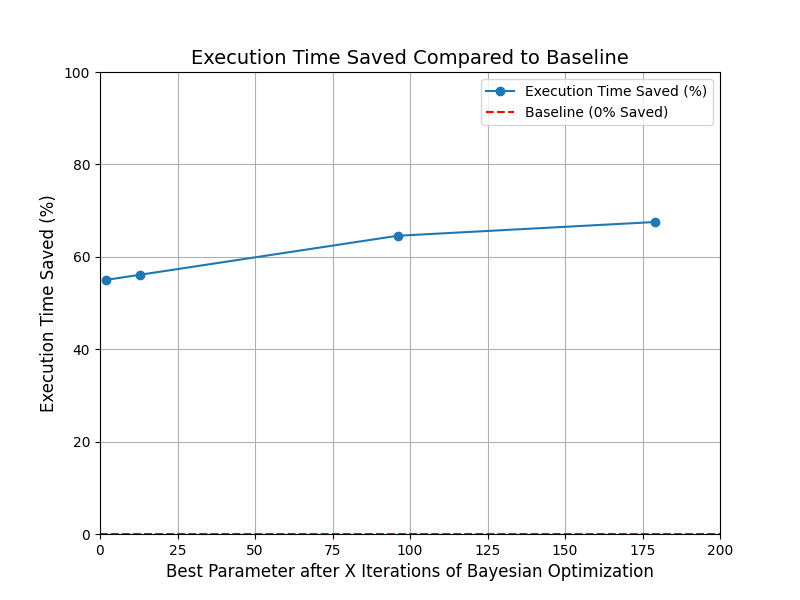}
    \caption{Percentage of Execution Time Saved by Bayesian ZebraConf Compared to Original ZebraConf at Different Iterations of Bayesian Optimization. Savings were 55\%, 56\%, 64\%, and 67\% for iterations $X = 2, 13, 96, 179$, respectively.}
    \label{fig:time_save}
\end{figure}

To evaluate execution time, we compare Bayesian-ZebraConf against ZebraConf when processing 1,000 unit test files. Figure~\ref{fig:test_time} shows the absolute execution times for \projname{} across iterations $X = 2, 13, 96, 179$ and ZebraConf, while Figure~\ref{fig:time_save} illustrates the percentage of time saved by \projname{} relative to ZebraConf.

\noindent \textbf{Observations.} 
Bayesian-ZebraConf consistently reduces execution time compared to ZebraConf. At iteration $X = 179$, \projname{} achieves a 67\% reduction in execution time, requiring only 153,417 seconds compared to ZebraConf's 472,503 seconds. These results demonstrate that \projname{} significantly accelerates testing by dynamically adapting hyperparameters and prioritizing high-risk configurations.

\noindent \textbf{Analysis.} 
The consistent improvements over iterations highlight the effectiveness of Bayesian optimization in refining the hyperparameters used in \projname{}. Earlier iterations ($X = 2, 13$) show modest gains due to the limited exploration of the hyperparameter space, while later iterations ($X = 96, 179$) stabilize with diminishing returns, indicating convergence towards optimal hyperparameters.

\subsection{Executed Test Count}

\begin{figure}[htbp]
    \centering
    \includegraphics[width=0.75\linewidth]{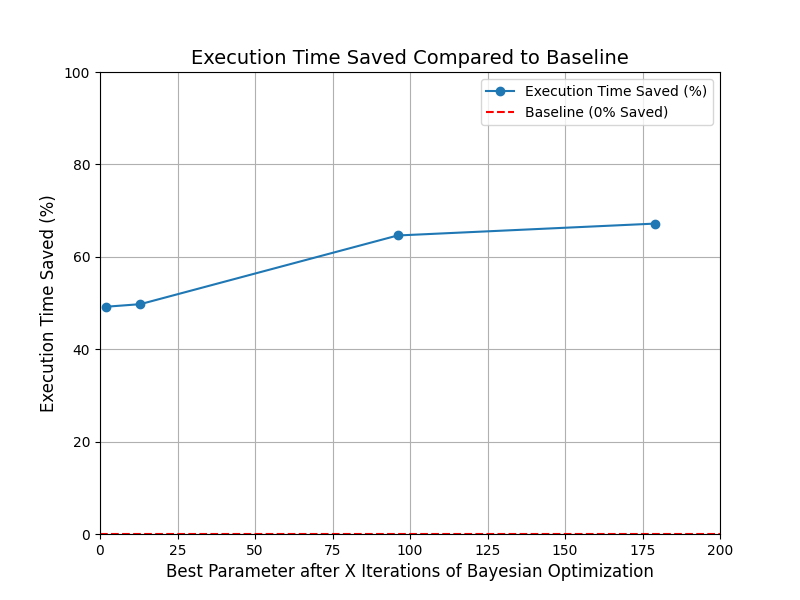}
    \caption{Comparison of the Number of Tests Executed by Bayesian ZebraConf (at Iterations $X = 2, 13, 96, 179$) and Original ZebraConf. Bayesian ZebraConf executed 54,850, 54,253, 38,204, and 35,426 tests, compared to ZebraConf's 107,962 tests.}
    \label{fig:test_count}
\end{figure}

\begin{figure}[htbp]
    \centering
    \includegraphics[width=0.75\linewidth]{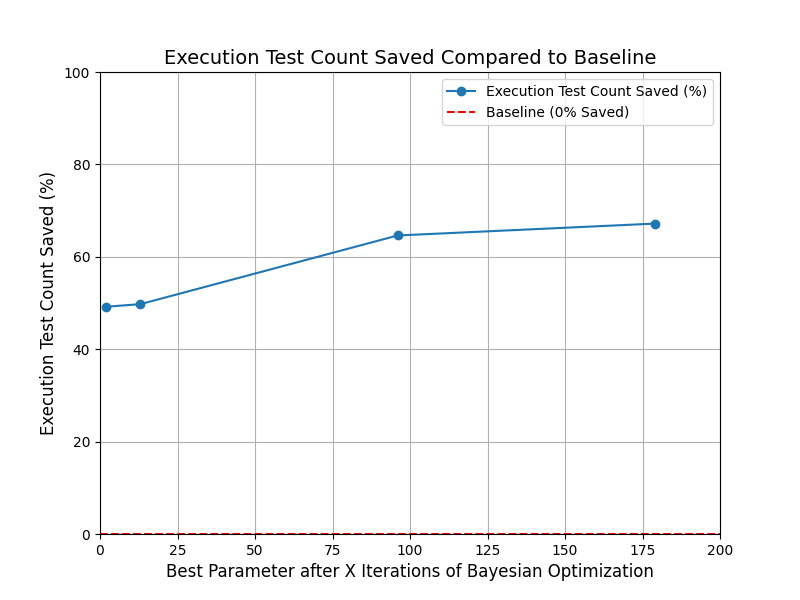}
    \caption{Percentage of Test Count Reduction Achieved by Bayesian ZebraConf Compared to Original ZebraConf Across Iterations $X = 2, 13, 96, 179$, with reductions of 49\%, 49\%, 64\%, and 67\%, respectively.}
    \label{fig:count_save}
\end{figure}

The efficiency of \projname{} is further evaluated by comparing the number of tests executed to classify 1,000 unit files. Figures~\ref{fig:test_count} and~\ref{fig:count_save} present the executed test counts and percentage reductions, respectively.

\noindent \textbf{Observations.} 
Bayesian ZebraConf reduces the executed test count from 107,962 (ZebraConf) to 35,426 at iteration $X = 179$, achieving a 67\% reduction. Similar to execution time, efficiency improves consistently across iterations, stabilizing at later stages.

\noindent \textbf{Analysis.} 
The reduction in executed tests directly correlates with the improved probabilistic modeling in BGT and dynamic adaptation of hyperparameters in BO. By prioritizing high-risk configurations, \projname{} minimizes redundant tests while maintaining accuracy, as demonstrated in the subsequent section on false positives and negatives.

\subsection{False Positive and False Negative Rates}

\begin{figure}[htbp]
    \centering
    \includegraphics[width=0.75\linewidth]{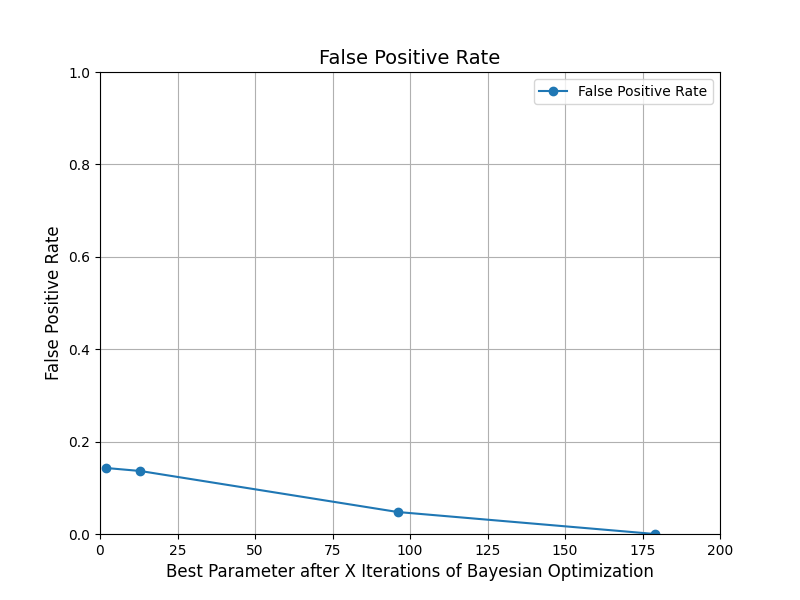}
    \caption{False Positive Rates of Bayesian ZebraConf Across Iterations $X = 2, 13, 96, 179$, with Rates of 14\%, 13\%, 4\%, and 0\%, respectively.}
    \label{fig:false_positive}
\end{figure}

\begin{figure}[htbp]
    \centering
    \includegraphics[width=0.75\linewidth]{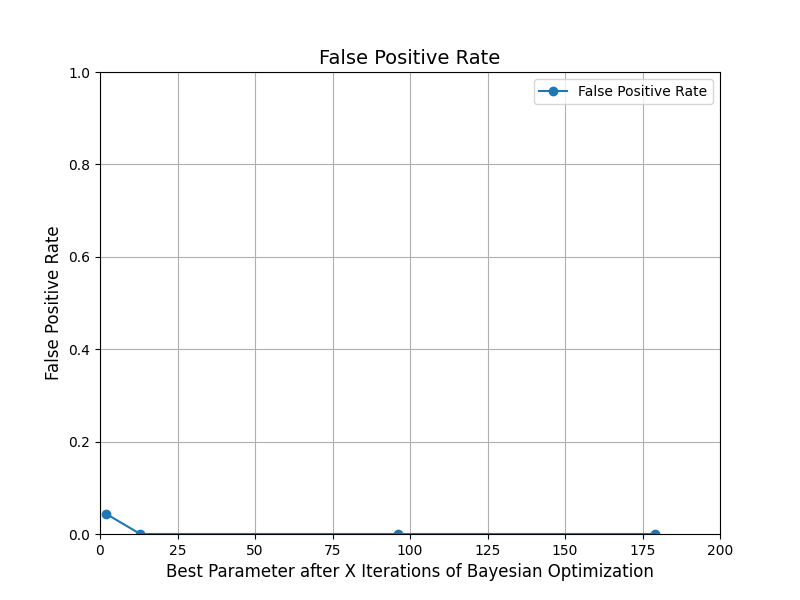}
    \caption{False Negative Rates of Bayesian ZebraConf Across Iterations $X = 2, 13, 96, 179$, with Rates of 4\% at $X = 2$ and 0\% for Subsequent Iterations.}
    \label{fig:false_negative}
\end{figure}

Figures~\ref{fig:false_positive} and~\ref{fig:false_negative} illustrate the false positive and false negative rates of \projname{} at different iterations.

\noindent \textbf{Observations.} 
The false positive rate decreases significantly, dropping to 0\% by iteration $X = 179$. False negatives are reduced to 0\% after the initial iteration.

\noindent \textbf{Analysis.} 
These improvements result from the adaptive thresholding mechanism in SBGT and the tuning of likelihood functions in BRR through BO. Stricter thresholds reduce misclassification rates but require more tests, highlighting the effectiveness of BO in balancing accuracy and efficiency.

\subsection{Insights and Additional Observations}

\noindent \textbf{Hyperparameter Optimization.} 
Bayesian optimization plays a pivotal role in achieving these results by dynamically tuning parameters such as pool sizes, prior probabilities, and classification thresholds. This iterative refinement ensures that the testing strategy adapts to evolving system behaviors.

\noindent \textbf{Probabilistic Reasoning.} 
The integration of probabilistic models in BGT and BRR allows \projname{} to prioritize high-risk configurations, reducing the number of redundant tests while maintaining accuracy.

\noindent \textbf{Convergence Trends.} 
The stabilization of improvements in later iterations indicates that \projname{} converges towards optimal performance, providing reliable and reproducible results.

\noindent \textbf{Future Improvements.} 
Exploring alternative acquisition functions in BO and hybrid group testing methods could further enhance \projname{}'s performance in diverse scenarios.

%% file: texts/conclusion.tex
\section{Conclusion}

Ba-ZebraConf significantly advances distributed system configuration analysis by addressing key limitations of existing frameworks like ZebraConf. By integrating Bayesian Group Testing (BGT), Bayesian Risk Refinement (BRR), and Bayesian Optimization (BO), it introduces a probabilistic, adaptive, and scalable approach to troubleshooting heterogeneous-unsafe configurations. Our evaluation demonstrates substantial improvements, with reductions in test execution time and test count by up to 67\%, while achieving 0\% false negative rates and eliminating false positives in advanced iterations. These results showcase the framework’s ability to dynamically prioritize high-risk configurations, accumulate evidence for accurate classification, and efficiently optimize hyperparameters through iterative refinement. The synergistic combination of BGT’s probabilistic inference, BRR’s cumulative risk assessment, and BO’s dynamic tuning underpin these advancements, ensuring robustness and efficiency in noisy and large-scale environments. 
Moving forward, we aim to explore alternative group testing methods to further enhance performance and extend the framework’s applicability to real-time and edge computing scenarios. By addressing these directions, Ba-ZebraConf has the potential to further solidify its role as a robust solution for next-generation distributed systems.